\documentclass[12pt]{article}
\usepackage{epsf}
\def\be{\begin{equation}}
\def\ee{\end{equation}}
\def\bea{\begin{eqnarray}}
\def\eea{\end{eqnarray}}

\begin{document}
\begin{titlepage}
\begin{center}
{\Large \bf William I. Fine Theoretical Physics Institute \\
University of Minnesota \\}
\end{center}
\vspace{0.2in}
\begin{flushright}
FTPI-MINN-15/17 \\
UMN-TH-3430/15 \\
April 2015 \\
\end{flushright}
\vspace{0.3in}
\begin{center}
{\Large \bf The process $e^+e^- \to \pi \pi X(3823)$ in the soft pion limit
\\}
\vspace{0.2in}
{\bf M.B. Voloshin  \\ }
William I. Fine Theoretical Physics Institute, University of
Minnesota,\\ Minneapolis, MN 55455, USA \\
School of Physics and Astronomy, University of Minnesota, Minneapolis, MN 55455, USA \\ and \\
Institute of Theoretical and Experimental Physics, Moscow, 117218, Russia
\\[0.2in]

\end{center}

\vspace{0.2in}

\begin{abstract}
The production of the resonance $X(3823)$, identified as the charmonium $^3D_2$ state,  in the process $e^+e^- \to \pi \pi X(3823)$ has been recently reported by BESIII. It is pointed out that this process is fully described, up to one overall coupling constant, in the soft pion limit. An interpretation of the available and possible future data within the discussed theoretical framework may reveal new features of the charmoniumlike states. In particular, the observed relative yield for this process at different energies strongly suggests a very significant enhancement of the amplitude at the charmoniumlike peak near 4.36\,GeV.     
\end{abstract}
\end{titlepage}

The charmonium state $X(3823)$ first observed~\cite{belle} through its radiative decay to $\gamma \chi_{c1}$ in the $B$ decays (and possibly also seen much earlier~\cite{e705} at a lower confidence level by the E705 experiment) is identified as the $1^3D_2$ charmonium level with the quantum numbers $J^{PC}=2^{--}$. Most recently the production of the same resonance in $e^+e^-$ annihilation was observed by BESIII~\cite{bes} through the process $e^+e^- \to \pi \pi X(3823)$ with a statistically significant yield at $\sqrt{s} = 4.42\,$GeV and  an indication of a comparable cross section at $4.36\,$GeV, although with a lower statistical significance due to a  smaller acquired integrated luminosity at the latter energy. The Ref.~\cite{bes} also presented the measured distribution in the dipion invariant mass and attempted an angular analysis under the assumption that `the $\pi^+ \pi^-$ system is very likely to be dominated by $S$-wave' (in the c.m. frame of the dipion). 

The purpose of the present paper is to point out that in as much as a soft pion approximation can be applied to the process $e^+e^- \to \pi \pi X(3823)$, the chiral symmetry uniquely determines, up to an overall constant, the behavior of the amplitude in the second order in the pion momenta including the effects of the symmetry breaking by the pion mass. This property does not rely on any additional assumptions besides the general chiral theorems, e.g. it is independent of the assumption of the heavy quark spin symmetry (HQSS) which is known~\cite{bes_h,lv} to be violated in similar processes in the same range of energy in the $e^+e^-$ annihilation~\footnote{The chiral theorems combined with HQSS were used~\cite{mv12} for description of similar processes in bottomonium, $e^+e^- \to \pi \pi ^3D_J$, for $J=1$, 2 and 3. In fact, as discussed here, it is sufficient to take into account only the chiral symmetry and its breaking to the first order in $m_\pi^2$ in order to fully describe the amplitudes for $J=2$ (as well as $J=3$) up to overall normalization constant(s) in the second order in the pion 4-momenta. Adding the arguments from HQSS allows to extend the description to the case of $J=1$ and to relate the normalization constants for all three values of $J$, which may not be justified for the discussed charmonium processes due to apparent violation of HQSS.}. 

In particular, the chiral theorems fully describe the shape of the spectrum in the dipion invariant mass as well as the dipion composition in terms of $S$- and $D$- waves in its c.m. frame. The latter composition significantly depends on the invariant mass, so that it would be impossible to describe the $\pi \pi$ system as `dominated' by one of the two waves in the entire phase space. 
Furthermore,  the amplitude of  $e^+e^-  \to \pi \pi X(3823)$ rapidly grows with the momenta of the pions. As a result the kinematical integral over the phase space is larger at $\sqrt{s} = 4.42\,$GeV than at $\sqrt{s} = 4.36\,$GeV by a factor of about three. Since the data~\cite{bes} indicate that the cross section at these energies is essentially the same, within the errors, one should conclude that the coupling at 4.36\,GeV is significantly stronger than at 4.42\,GeV. In other words, the observed process is likely due to the peak $Y(4360)$ which is known to be present in the channels $e^+e^- \to \pi \pi \, \psi(2S)$~\cite{belle436,lqy} and $e^+e^- \to \pi \pi \, h_c$~\cite{bes_h}. This conclusion also certainly agrees with the observed cross section for $e^+e^-  \to \pi \pi X(3823)$ being compatible with zero at all the reported energies, both low and high, except for 4.36\,GeV and 4.42\,GeV.

In order to apply the chiral symmetry requirements to the process $e^+e^-  \to \pi \pi X(3823)$  one can notice that its amplitude can uniquely be written in the form
\be
A \left[ e^+ e^- \to \pi \pi X(3823) \right ] = \epsilon_{\mu \nu \lambda \sigma} \, F_{\mu \nu} \, \psi_{\lambda \kappa} \, T_{\sigma \kappa} (p_1, p_2)~,
\label{gena}
\ee
where $\epsilon_{\mu \nu \lambda \sigma}$ is the antisymmetric simbol, $F_{\mu \nu}$ is the field strength tensor for the virtual photon, $\psi_{\lambda \kappa}$ is the symmetric spin-2 wave function of the resonance $X(3823)$, and $T_{\sigma \kappa} (p_1, p_2)$ is a symmetric tensor depending on the 4-momenta $p_1$ and $p_2$ of the pions. Due to the Bose symmetry the latter tensor has to be symmetric under the interchange of the pions, $p_1 \leftrightarrow p_2$, and the chiral symmetry requires that it is vanishing when one of the pion 4-momenta goes to zero (with all other particles, including the other pion, being on mass shell). Clearly there is only one structure in $T_{\sigma \kappa}$ in the second order in the pion momenta that satisfies this condition and gives a nonvanishing contribution in the amplitude (\ref{gena}): $T_{\sigma \kappa} (p_1, p_2) = C \, (p_{1 \sigma} p_{2 \kappa} + p_{1 \kappa} p_{2 \sigma})$ with $C$ being a constant. It should be noted that this form is also valid if the violation of the chiral symmetry by the pion mass, $m_\pi^2$, is taken into account. Indeed, the term in $T_{\sigma \kappa}$ proportional to $m_\pi^2$ (i.e. also of the second order in the pion momenta) could only enter being multiplied by the metric $g_{\sigma \kappa}$. (More specifically, the term, satisfying the Adler zero condition is proportional to $g_{\sigma \kappa} \, [ (p_1+p_2)^2 - m_\pi^2]$.) However, the contribution of such term in the amplitude (\ref{gena}) would be zero due to the symmetry of $\psi_{\lambda \kappa}$.  As a result the general quadratic in the pion momenta expression for the amplitude can be written, up to an overall constant $C_1$, as
\be
A \left[ e^+ e^- \to \pi \pi X(3823) \right ] = C_1 n_\mu \, \epsilon_{\mu \nu \lambda \sigma} \, j_\nu \, \psi_{\lambda \kappa} \, (p_{1 \sigma} p_{2 \kappa} + p_{1 \kappa} p_{2 \sigma}) = - C_1 \, \epsilon_{nls} \, j_n \, \psi_{lk} \, (p_{1s} p_{2k} + p_{1k} p_{2 s})~,
\label{amp}
\ee
where $n_\mu$ is a unit vector with components $(1,0,0,0)$ in the c.m. frame of the colliding $e^+$ and $e^-$ beams, $j_\nu$ is the electromagnetic current of the colliding particles (so that $F_{\mu \nu} \propto n_\mu j_\nu - n_\nu j_\mu$), and the latter expression in Eq.(\ref{amp}) is written in the the same frame in terms of only the relevant spatial components. The middle expression is helpful in transforming to the c.m. frame of the dipion and separating the $S$-wave and $D$-wave in that frame, while the latter expression is useful for describing the correlations in the c.m. frame of the beams, e.g. between the directions of the pions and the beam axis, which correlations are discussed in Ref.~\cite{mv12}.

Here we concentrate on the contribution of the two partial waves for the dipion and the distribution in its invariant mass. Defining $q=p_1+p_2$ and $r=p_1 - p_2$, and also the spin-2 tensor
\be
\ell_{\sigma \kappa} = r_\sigma r_\kappa - {1 \over 3} \, \left ( 1 - {4 \, m_\pi^2 \over q^2} \right ) \, \left ( q^2 \, g_{\sigma \kappa}- q_\sigma  q_\kappa \right )~,
\label{eldef}
\ee
the middle expression in Eq.(\ref{amp}) can be written as
\be
A \left[ e^+ e^- \to \pi \pi X(3823) \right ] = {C_1 \over 2} \, n_\mu \, \epsilon_{\mu \nu \lambda \sigma} \, j_\nu \, \psi_{\lambda \kappa} \, \left [ {2 \over 3} \, \left ( 1+ {2 \, m_\pi^2 \over q^2} \right ) \, q_\sigma q_\kappa - \ell_{\sigma \kappa} \right ]~.
\label{sdamp}
\ee
This expression explicitly separates the dipion $S$-wave and $D$-wave states: the term with $q_\sigma q_\kappa$ describes the $S$-wave, while that with $\ell_{\sigma \kappa}$ is the $D$-wave term. One can readily see that in this expression the relation between the two waves is rigidly fixed by the chiral symmetry. Accordingly, if one writes the expression for the distribution of the rate in the invariant mass $m_{\pi \pi}$ ($q^2 = m^2_{\pi \pi}$) as the sum of the $S$- and $D$-wave contributions
$d \sigma / d m_{\pi \pi} = C_2 \, (\rho_S + \rho_D)$ with a common overall constant $C_2$, the expressions for the spectral densities $\rho_S$ and $\rho_D$ read as
\bea
&& \rho_S= {8 \over 9} \, \sqrt{q^2-4\, m_\pi^2} \, (\Delta^2 - q^2)^{5/2} \, \left (1+ {2 \, m_\pi^2 \over q^2} \right )^2~, \nonumber \\
&& \rho_D = 2 \sqrt{q^2-4\, m_\pi^2} \, \sqrt{\Delta^2 - q^2} \, \left (1 - {4 \, m_\pi^2 \over q^2} \right )^2 \, \left [ (q^2)^2 + {2 \over 3} \, q^2 (\Delta^2-q^2) + {4 \over 45} \,  (\Delta^2-q^2)^2 \right ]~,
\label{sdd}
\eea
where the recoil of the heavy state $X(3823)$ is neglected, so that $\Delta = \sqrt{s} - M_X$ is the total energy of the two pions, $\Delta = (n \cdot q)$.

\begin{figure}[ht]
\begin{center}
 \leavevmode
    \epsfxsize=11cm
    \epsfbox{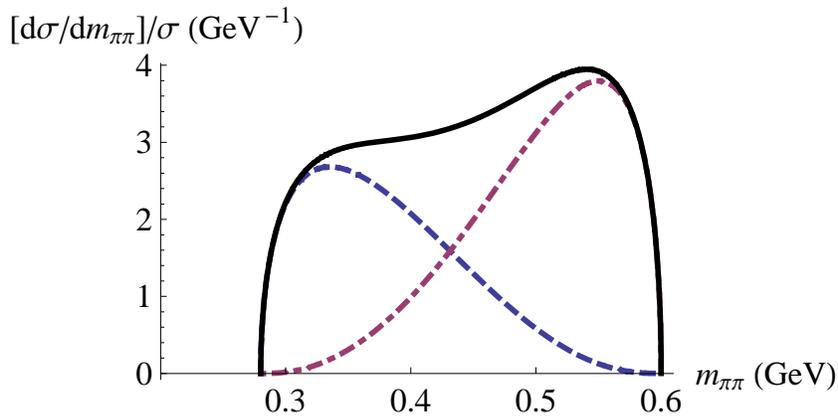}
    \caption{The contribution of the production of the $S$-wave (dashed) and the $D$-wave (dotdashed) $\pi \pi$ system to the rate of the process $e^+e^-  \to \pi \pi X(3823)$ at $\sqrt{s}=4.42$\,GeV. Also shown is the distribution of the total rate (solid).}
\end{center}
\end{figure} 

The dependence on $m_{\pi \pi}$ of the contribution of the $S$- and $D$- wave dipion production to the rate as well as the total distribution in $m_{\pi \pi}$ described by the expressions (\ref{sdd}) is illustrated in Fig.~1 (at the total energy $E=\sqrt{s}=4.42$\,GeV, chosen for definiteness). One can readily see that the relative importance of the two partial waves very substantially depends on $m_{\pi \pi}$, so that it is impossible to consider the process as being dominated by one or the other wave over the whole kinematical range. 

The number of observed events in Ref.~\cite{bes} is too small to make a meaningful comparison with the distribution in $m_{\pi \pi}$ described by Eq.(\ref{sdd}). If future data indicate a significant deviation from the soft-pion description, this would imply a presence of a nontrivial dynamics in one or more channels. Should this be the case, it would be unlikely that the chiral expansion is invalidated due to the dynamics in the $\pi \pi$ channel (the $\pi \pi$ rescattering), since this would invalidate the agreement of the soft pion description with the data in other processes where it is well established, e.g. in the transitions $\psi(2S) \to \pi \pi \, J/\psi$ and $\Upsilon(2S) \to \pi \pi \, \Upsilon(1S)$~\cite{bc,mv75} (for a review see e.g. Ref.~\cite{mv_ch}). In those processes the soft pion expansion agrees with the data up to the available energy $\Delta \approx 0.6$\,GeV, which in terms of the discussed process $e^+e^-  \to \pi \pi X(3823)$ corresponds to the c.m. energy up to 4.42\,GeV. The chiral expansion may be invalidated if the process is dominated by a presently unknown isovector charmoniumlike relatively narrow resonance $Z$ with mass near the kinematical boundary, so that the production of $X(3823)$ goes in two stages: $e^+e^- \to \pi Z \to \pi \pi X(3823)$ with one very soft pion being emitted in the $S$-wave (in the c.m. frame of the initial beams) and the other, more energetic, in the $D$-wave. Clearly, this mechanism implies that the mass of $Z$ is either near 4220\,MeV or near 3965\,MeV. In the former case the quantum numbers of $Z$ should be $I^G(J^P)=1^+(1^+)$ [similar to the most plausible assignment for $Z_c(3900)$ and $Z_c(4020)$], while in the latter case these numbers should be $1^+(2^+)$, i.e. totally different from any hadron known so far. Although this possibility is presently rather remote, it may still justify an effort towards further experimental study of the discussed process.

It can be also noticed that the total phase space integral over the spectral factors $\rho_S$ and $\rho_D$:
\be
\Phi(E)=\int_{2 m_\pi}^{E-M_X} \, \left (\rho_S + \rho_D \right ) \, d m_{\pi \pi}~,
\label{fi}
\ee
very rapidly grows with the available energy $\Delta$ for the pions. The behavior of this integral is shown in Fig.~2. This growth however is entirely due to the kinematics and the soft pion dynamics required by the chiral symmetry, rather than being related to the dynamics of the heavy charmed quarks. The intrinsic strength of the coupling of the hidden charm states produced in the $e^+e^-$ annihilation to the final state $\pi \pi X(3823)$ is characterized by the coupling $C$. According to the data~\cite{bes} the cross section measured at 4.36\,GeV and 4.42\,GeV is approximately the same. Taking into account the ratio of the phase space integrals, $\Phi(4.42\,$GeV$)/\Phi(4.36\,$GeV$) \approx 2.8$, one should conclude that the coupling of the hidden charm state at 4.36\,GeV  
is significantly stronger than at 4.42\,GeV.

\begin{figure}[ht]
\begin{center}
 \leavevmode
    \epsfxsize=11cm
    \epsfbox{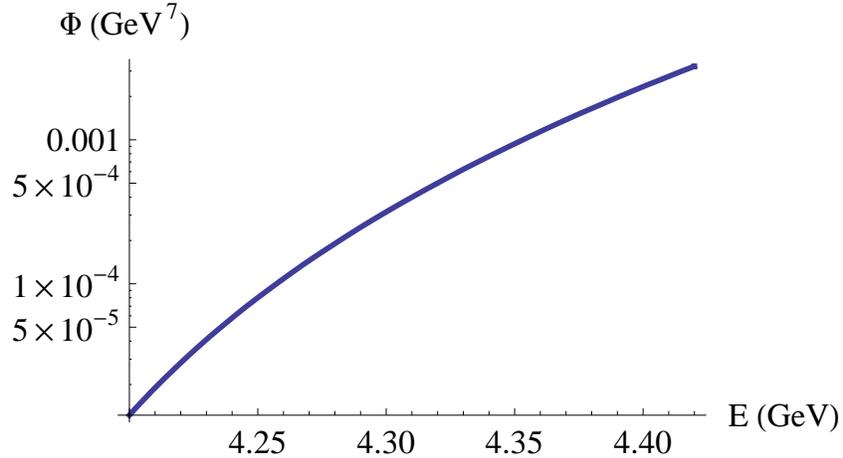}
    \caption{The dependence of the phase space integral $\Phi$ on the total c.m. energy.}
\end{center}
\end{figure} 

Certainly, the conclusion about the relative strength of the coupling $C$ for the hidden charm states with different mass would significantly change, if a production of $\pi \pi X(3823)$ is measured at lower energy $E$ where the phase space factor $\Phi$ is much smaller. The presently available data at $E=4.23\,$GeV and $4.26$\,GeV indicate a signal with a statistical significance of less than 2$\sigma$, so that the status of the discussed process at those energies is yet to be clarified. Similarly uncertain is the status of the data at the highest measured energy 4.6\,GeV, where the factor $\Phi$ calculated from Eq.(\ref{sdd}) is very large. It should be noted however that at such energy there could be noticeable deviations from the soft pion expansion for the amplitude of the discussed process. 

Before concluding this paper it is worth mentioning that the charmonium fine-structure partners of the $X(3823)$ resonance, the $^3D_1$ [identified with $\psi(3770)$] and the $^3D_3$ (yet unknown) states, can also be produced in a similar process in the electron-positron annihilation, $e^+e^- \to \pi \pi \, ^3D_J$~\footnote{For those states however the decay into the $D \bar D$ pairs is allowed by the quantum numbers, so that the branching fraction for their radiative transitions to the $\chi_{cJ}$ levels is small, and they are not visible in the experimental setting reported in Ref.~\cite{bes}.}. In fact using the HQSS the processes for all three values of $J$ are related~\cite{mv12} in the soft pion limit~\footnote{I take this opportunity to correct the previous erroneous remark~\cite{mv12} regarding the relative yield of the $^3D_J$ states in the limit of exact HQSS degeneracy being proportional to $2J+1$. The correct ratio is 1:3:5 for $J=1$, 2 and 3.}. However, beyond the HQSS approximation such relation is generally lost. In particular, the amplitude for the production of the $^3D_1$ state can generally receive an additional part with the emission of the $\pi \pi$ system in the $S$-wave. (An example of mechanism for such contribution is e.g. the $^3S_1 - ^3D_1$ mixing.) Nevertheless, the structure of the production amplitude for the spin-3 $^3D_3$ state is uniquely fixed up to an overall coupling constant, in the same way as discussed here for the spin-2 $X(3823)$ state. Namely, at a zero recoil the amplitude in the c.m. frame can be written similarly to the latter expression in Eq.(\ref{amp}) as
\be
A(e^+e^- \to \pi \pi \, ^3D_3) = C_3 \, j_i \, \psi_{ikl} \, (p_{1k} p_{2l} + p_{1l} p_{2k} )~,
\label{amp3}
\ee
in terms of the spin-3 polarization wave function $\psi_{ikl}$ (totally symmetric and traceless). This structure of the amplitude results in exactly the same formulas as in Eq.(\ref{sdd}) for the distribution of the production of $S$-wave and $D$-wave $\pi \pi$ system over the invariant mass $m_{\pi \pi}$, and the angular correlations in the process can be found in Ref.~\cite{mv12}. 

This work  is supported, in part, by U.S. Department of Energy Grant No. DE-SC0011842.


\begin{thebibliography}{99}
\bibitem{belle} 
  V.~Bhardwaj {\it et al.}  [Belle Collaboration],
  %``Evidence of a new narrow resonance decaying to $\chi_{c1}\gamma$ in $B \to \chi_{c1} \gamma K$,''
  Phys.\ Rev.\ Lett.\  {\bf 111}, no. 3, 032001 (2013)
  [arXiv:1304.3975 [hep-ex]].
  
\bibitem{e705} 
  L.~Antoniazzi {\it et al.}  [E705 Collaboration],
  %``Search for hidden charm resonance states decaying into J / Psi or Psi-prime plus pions,''
  Phys.\ Rev.\ D {\bf 50}, 4258 (1994).
  
\bibitem{bes} 
  M.~Ablikim {\it et al.}  [BESIII Collaboration],
  %``Observation of the $\psi(1^3D_2)$ state in $e^+e^-\to\pi^+\pi^-\gamma\chi_{c1}$ at BESIII,''
  arXiv:1503.08203 [hep-ex].
  
\bibitem{bes_h} 
  M.~Ablikim {\it et al.}  [BESIII Collaboration],
  %``Observation of a Charged Charmoniumlike Structure $Z_c$(4020) and Search for the $Z_c$(3900) in $e^+e^- \to ?^+?^-h_c$,''
  Phys.\ Rev.\ Lett.\  {\bf 111}, no. 24, 242001 (2013)
  [arXiv:1309.1896 [hep-ex]].
\bibitem{lv} 
  X.~Li and M.~B.~Voloshin,
  %``$Y$(4260) and $Y$(4360) as mixed hadrocharmonium,''
  Mod.\ Phys.\ Lett.\ A {\bf 29}, no. 12, 1450060 (2014)
  [arXiv:1309.1681 [hep-ph]].
 
\bibitem{mv12} 
  M.~B.~Voloshin,
  %``Heavy quark spin symmetry breaking in near-threshold $J^{PC}=1^{--}$ quarkonium-like resonances,''
  Phys.\ Rev.\ D {\bf 85}, 034024 (2012)
  [arXiv:1201.1222 [hep-ph]].
  
\bibitem{belle436} 
  X.~L.~Wang {\it et al.}  [Belle Collaboration],
  %``Observation of Two Resonant Structures in e+e- to pi+ pi- psi(2S) via Initial State Radiation at Belle,''
  Phys.\ Rev.\ Lett.\  {\bf 99}, 142002 (2007)
  [arXiv:0707.3699 [hep-ex]].
  
\bibitem{lqy} 
  Z.~Q.~Liu, X.~S.~Qin and C.~Z.~Yuan,
  %``Combined fit to BaBar and Belle data on e+e- ---> pi+ pi- psi(2S),''
  Phys.\ Rev.\ D {\bf 78}, 014032 (2008)
  [arXiv:0805.3560 [hep-ex]].  

\bibitem{bc} 
  L.~S.~Brown and R.~N.~Cahn,
  %``Chiral Symmetry and psi-prime ---> psi + pi + pi Decay,''
  Phys.\ Rev.\ Lett.\  {\bf 35}, 1 (1975).

\bibitem{mv75} 
  M.~B.~Voloshin,
  %``Adler's Selfconsistency Condition in the Decay psi-prime (3700) --> psi (3100) pi pi,''
  JETP Lett.\  {\bf 21}, 347 (1975)
  [Pisma Zh.\ Eksp.\ Teor.\ Fiz.\  {\bf 21}, 733 (1975)].
  
  
\bibitem{mv_ch} 
  M.~B.~Voloshin,
  %``Charmonium,''
  Prog.\ Part.\ Nucl.\ Phys.\  {\bf 61}, 455 (2008)
  [arXiv:0711.4556 [hep-ph]].  

\end{thebibliography}
\end{document}